\documentclass[11pt,twoside]{article}

\usepackage{asp2006}
\usepackage{epsf}
\usepackage{psfig}
\usepackage{lscape}
\usepackage{graphicx}

\begin{document}
\title{Meet the COG's}  
\author{Martin Altmann$^1$, M\'arcio Catelan$^2$, Manuela Zoccali$^2$} 
\affil{$^1$Astronomisches Recheninstitut, Universit\"at Heidelberg, M\"onchhofstrasse 12-14, 69120 Heidelberg, Germany\\
$^2$Pontificia Universidad Cat\'olica de Chile, Departamento de Astronom\'ia y Astrof\'isica, Av. Vicu\~na Mackenna 4860, 782-0436 Macul, Santiago, Chile}  

\begin{abstract} 
Kinematics combined with detailed element abundances provide a method
of analysis of stellar populations that uses as much available information 
as possible, in contrast to other methods. Here we employ this technique on 
local A-type horizontal branch stars in an ongoing programme to search 
for $\omega$ Centauri debris among these objects. This led us to discover
another group of stars with very similar kinematics and abundances, the 
{\bf C}ometary {\bf O}rbit {\bf G}roup (COG). We also comment on the future 
of this kind of undertaking in the Gaia era. 
\end{abstract}

\section{Introduction: Studying Galactic Structure}
There are quite a few approaches when it comes to the study of the stellar
component of our Galaxy. Each of them has its merits and caveats, and in
an overall picture these different methods complement each other. 
There are several important items to consider when planning a project
aiming at analysing Galactic structure: 
\begin{itemize}
\item What part of the Galaxy is the aim? Is it young or old? What has 
previously been known about it? How can previous results be improved on 
by adding new data or better defined and larger samples?  
\item What is the best approach to achieve the goal, i.e. is it better to
have a large sample at the expense of not having all information and/or
tolerating less accurate information, or study a limited sample with more
detailed and better information available? Obviously, it would be ideal to have 
as large a sample as possible with all information; however, in most cases 
this is not feasible. High-quality information such as (especially high-resolution) 
spectroscopy is expensive in terms of observing time and might even be beyond 
the reach of current telescopes for faint objects. The fortunate situation 
today is that more and more ambitious programs have become feasible.
\item Selection effects and statistical ambiguities: every sample, however carefully
composed, is subject to selection effects. Owing to the Sun's location within the 
Milky Way, the pitfalls of selection effects are nowhere so 
common as in Galactic astronomy. While many of these can be accounted for in a statistical
sense, some can only be described in a qualitative way, and need to be kept in mind when interpreting
the results. Minimising such unwanted systematic side effects is one of the most important constraints
when undertaking a study of Galactic structure.
\end{itemize}

Several classical methods exist, some of them having been in use for almost a century. The simplest
one is star counts, especially in pencil beam fields \citep[see e.g.][]{Altmann2007}. 
Here the spatial distribution of all stars
or stars of a given temperature or colour range in a well defined field in the sky is studied. For this
one needs to determine distances by means of photometric or spectroscopic parallaxes. Star counts can 
also be accomplished on the whole sky or parts of the sky, which allows for the detection of smaller-scale
anomalies such as streams and moving groups. 

\begin{figure}
\begin{center}
\includegraphics[scale=0.6]{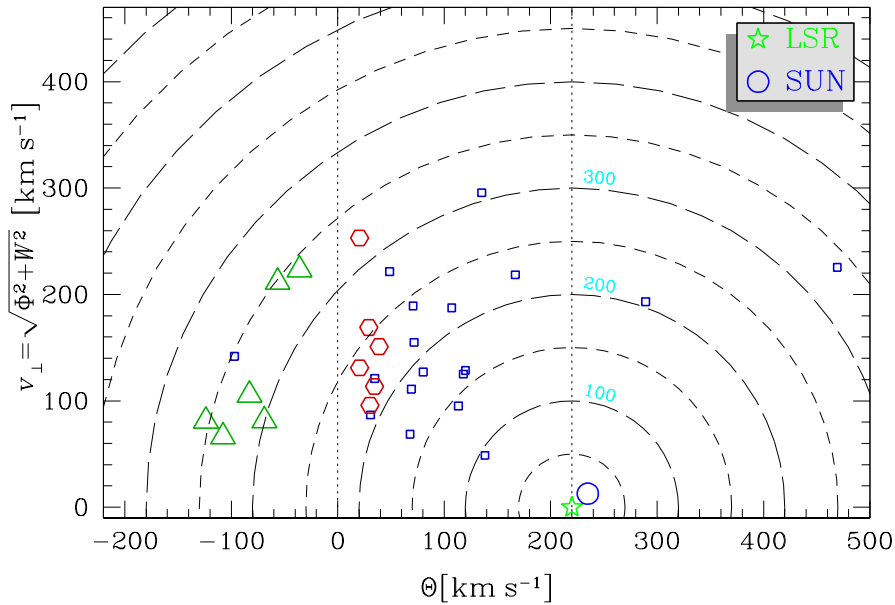}
\end{center}
\caption{Toomre diagram for the HBA stars in our sample \citep[Fig. 1 of][]{Altmann2005}.
The concentric circles
show the absolute peculiar velocity
($v_{\rm pec}$), i.e. the
total deviation from circular velocity (shown as an open star; the
Sun's position in this diagram is shown as an open circle).
The hexagons depict the stars belonging to the members of the
Cometary Orbit Group (COG) discussed in this
paper, the triangles denote putative $\omega$ Cen debris candidates (identified
as retrograde moving stars broadly conforming to the abundance range of $\omega$~Cen giants),
the other HBAs are represented by squares.}
\end{figure}
\vfill

Adding kinematics gives access to more parameters, such as the movement of objects. While a purely 
spatial analysis allows for the description of populations such as Galactic disks in a statistical
sense, knowledge of the kinematics allows to assign particular objects to stellar groups or populations. 
It would be ideal to have the complete kinematical information, i.e. radial velocities {\it and} proper 
motions, but significant progress has been made using one component alone, such as radial velocities in the
case of the discovery of the Sagittarius dwarf \citep{Ibata94}. Target objects can be stars in general or
specific tracers, like white dwarfs \citep[see, e.g.,][]{Pauli2003,Pauli2006} or horizontal branch 
(HB)-like stars.  For example, blue subdwarf (sdB) 
stars have been in the focus of several 
such studies \citep[see, e.g.,][]{deBoer1997,Altmann2004}. 
The study of \citet{Altmann2000} dealt with the complete HB blueward of the instability strip, 
while \citet{Kaempf2005} studied red HB (RHB) and \citet{Maintz2005} RR Lyrae stars \citep[see][for a 
description of the several HB components]{mc05}. 
Caveats of using kinematics include variability of radial velocities
caused by close binarity or pulsations. Proper motions are often difficult to find in the literature and take
a long time to obtain, since one needs data from two epochs with as large as possible a time baseline. 
In recent years, several whole-sky catalogues \citep[e.g.,][]{USNO} have appeared, improving
the situation dramatically. Space-based astrometry from Hipparcos \citep{Hipparcos} provided excellent
proper motions for about 120,000 stars. 

A further step in complexity is to add abundances to the kinematic analysis. It is known that the various
components of the Galaxy have different abundance patterns, i.e. ratios between various element groups,
such as iron peak and $\alpha$-capture elements \citep[see, e.g.,][and references therein]{Fuhrmann1998,mc07}. 
Differential abundances add
important information about memberships of stars to certain groups and also the evolution of the  
Galaxy as a whole, and thus of galaxies in general. Since every population of stellar objects has its distinct
star formation history as revealed by its abundance pattern, knowing kinematics {\it and} differential
abundances allows us to distinguish objects of different origin from within the Galaxy or from its outside, 
i.e. in dwarf galaxies that have been accreted by the Milky Way. 
A typical example for such a chemodynamical study is given by
\citet{Altmann2005}, which will also be the focus of the remainder of this paper. 

\section{Abundances and Kinematics of Local HBA Stars}
A-type HB (HBA) stars, in contrast to their hotter siblings---the B-type HB (HBB) and sdB/OB stars---have 
abundance patterns unaltered
by diffusion and levitation processes. Therefore, they are an ideal subject for such an undertaking. 
Presently we restrict ourselves to very local HBA stars, i.e. those well-known objects located within 
1~kpc of the Sun. These stars have a significant amount of data available, allowing them to be studied on 
the basis of archival and literature data alone. Our original goal was to carry out a preliminary study
in the quest for $\omega$~Centauri debris that might be present in the field.

\subsection{The peculiarities of $\omega$ Centauri}

$\omega$~Centauri (NGC~5139) 
is the most massive and brightest globular cluster of the Milky Way. In
contrast to most other globular clusters, this object contains more than one population of stars, 
with different ages and abundances. In contrast to most objects in the Galaxy, it is on a
retrograde orbit.
Its outstanding position with respect to ``normal'' globulars has led to the notion that it might in fact
be the nucleus of a smaller galaxy that at one time was incorporated into our own Galaxy. This was further
substantiated when the second most massive globular cluster, M\,54, was found to belong to the Sagittarius 
dwarf spheroidal, which is currently in the process of being swallowed by the Galaxy \citep{Layden2000}. 
\citet{Bekki2003} and \citet{Dinescu2002} have calculated the kinematics of possible debris, considering 
$\omega$~Cen as the remains of a collision event with subsequent accretion. Their results were then used
by us to determine whether an object of our sample is a viable $\omega$~Cen debris candidate or not. 

\begin{figure}
\begin{center}
\includegraphics[scale=0.5]{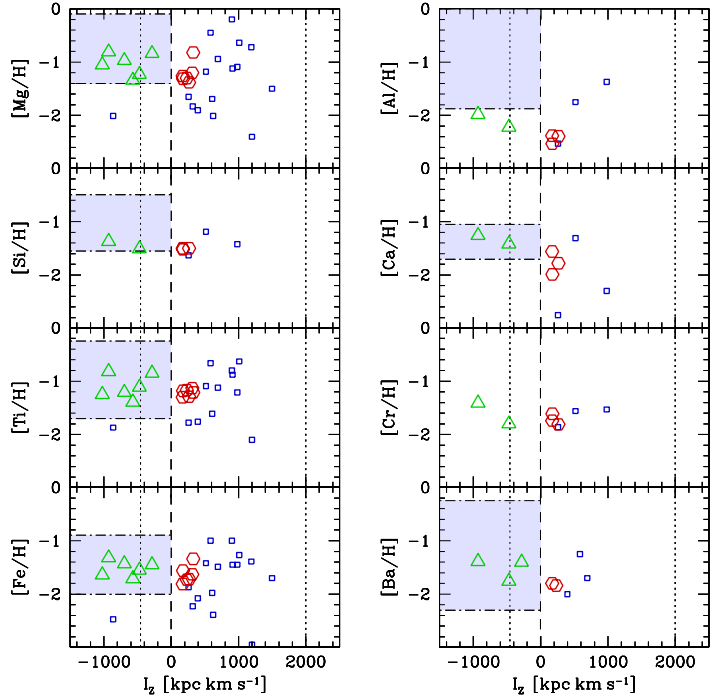}
\end{center}
\caption{[X/H] vs. $I_z$ diagram for eight different elements \citep[Fig. 2 of][]{Altmann2005}.
    The vertical dashed line marks the boundary
    between pro- and retrograde rotation, while the dotted lines
    $\omega$~Cen's $I_z$ (left) and that of the Sun (right) to represent the
    $I_z$ of a typical disk star. The shaded area between the horizontal lines in
    the retrograde part of the plots denotes the abundance
    range of each element in $\omega$~Cen, as conservatively derived from
    the extreme values of the relatively small sample of stars analysed by Smith et al.
    (2000).
    The object symbols are used as described in Fig. 1.
    Two extreme objects are outside the $I_z$ range of this plot.
}
\end{figure}
\vfill

The abundance pattern of $\omega$~Cen
is very peculiar, especially in the element species
O, Na, Mg, Cu, and other s-process elements \citep{Norris1995a,Norris1995b,Smith2000}. 
This would make every object formerly belonging to an entity of which $\omega$~Cen
was the nucleus stand out clearly in every sample of stars. Given that $\omega$~Cen contains a large
number of HBA stars (but largely lacks RHB stars), these objects are an excellent tracer for $\omega$~Cen 
debris in the field.

\subsection{Assembling the Sample}
Our sample of 30 HBA stars was assembled from archival and literature data,
most of the objects already being in the sample
of \citet{Altmann2000}. Proper motions were taken from the Hipparcos catalogue, radial velocities from various
sources in the literature, and abundances mostly from \citet{Kinman2000} and \citet{Behr2003}.
The assembly of the sample and the accumulation of data is described in more detail in \citet{Altmann2005}.
While we have data for eight elements, for several of them we only have data for a few stars. Therefore a follow-up
study with more uniform abundance data is mandatory.

\section{Results}

\begin{figure}
\begin{center}
\includegraphics[scale=0.50]{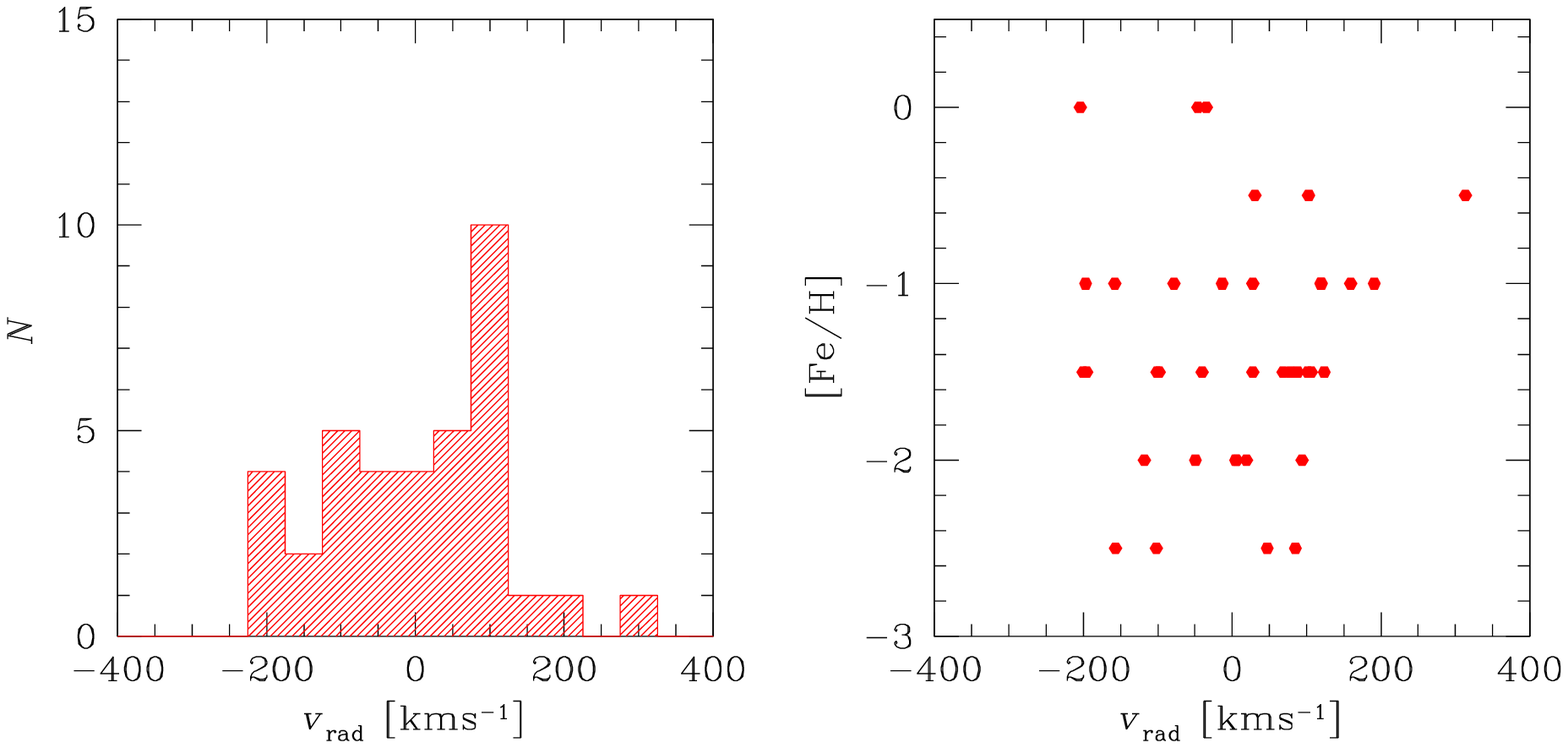}
\end{center}
\caption{The BHB stars of \citet{Peterson2001} (Figure adapted from \citet{Altmann2002}). Note the relatively strong peak at $v_{\rm rad}i= 100 $ km\,$s^{-1}$ in 
the left panel and the concentration of data points near $v_{\rm rad}i= 100 $ km\,s$^{-1}$ and [Fe/H]=-1.5 dex. Some of
the stars forming this overdensity might be related to the COGs. A more detailed study of the kinematic behaviour and
detailed abundances of the \citet[stars]{Peterson2001} will show whether there is any relationship to our COG stars.
}
\end{figure}

\subsection{$\omega$~Cen candidates}
Indeed, 7 of our 30 stars (or 23\%) are on retrograde orbits (see Figs 1 and 2), and from a kinematical point of view they could be
$\omega$~Cen debris. However, one of them (HD~87047) is far too metal-poor to have once been associated to 
$\omega$~Cen. The other six objects are still viable candidates; it is only with our current follow-up study 
based on or own spectroscopic data spanning the whole optical range that will determine which (if any) of 
these stars are really connected to $\omega$~Cen (see Fig. 2). 

\subsection{A Surprising Discovery: The {\bf C}ometary {\bf O}rbit {\bf G}roup} 
More exciting than the very preliminary results on whether our sample includes any candidate for $\omega$~Cen debris
was the truly unexpected discovery of a group of 6 stars with very closely related kinematics and abundances. These stars
are on slightly prograde orbits with a very low orbital velocity (see Figs 1 and 2); hence their orbits are cometary, leading them very
close to the Galactic centre. Therefore we dubbed this group of stars the {\bf C}ometary {\bf O}rbit {\bf G}roup, or
the COG's. The close proximity of the perigalacticon to the Galactic centre causes the orbits
to be wildly chaotic, which is shown in wildly different maximal $z$-heights. The apogalactic distances are 
for almost all stars between 8 and 10~kpc, except for HD~86986 which ventures to almost 17~kpc from the centre. 
This makes this object the least likely to be part of the COG.  

The abundance pattern\footnote{For those stars for which data are available.} is very similar in all of the 
analysed elements except Calcium, where the abundances of 3 stars show a bit more variation. Very striking
is the similarity in elements such as Fe, Mg and Ti (for which we have data for most of the stars), 
the abundances of which are virtually identical. The [Fe/H] abundance of the group is about $-1.6$ to 
$-1.7$~dex (see Fig. 2).

The nature of this group remains unknown. Given its very degraded orbital pattern, the object of which the COGs are
the remains must have been accreted into the Galaxy a rather long time ago. Since these objects are on such chaotic
orbits the COGs should not be restricted to the solar neighbourhood but exist in rather large parts of the 
Galaxy. These stars may have originated during the initial collapse of the proto-Galaxy that may have led to 
the formation of the present-day disk system, a process originally described in \citet{ELS1962}. 
More evidence for such an origin might be given by the overdensity of stars with an [Fe/H] abundance 
of $\sim-1.5$ to $-2$~dex in the bulge sample of \citet{Peterson2001}, as found by \citet{Altmann2002}.
Unfortunately, the abundances of \citeauthor{Peterson2001} are only good to 0.5~dex, 
and they do not have proper motions for their objects (see Fig. 3).

\begin{figure}[t!]
\begin{center}
\includegraphics[scale=0.4]{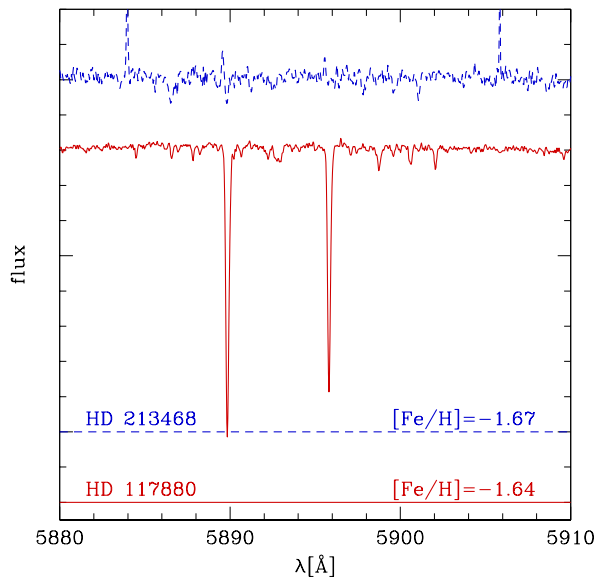}
\end{center}
\caption{Comparison of the Na D lines of two stars which are otherwise very similar in temperature, $\log g$
and [Fe/H].
While HD 117880 has very prominent Na D dublett, HD 213468 has almost no trace of Na lines. Closer inspection
of the spectrum (or this figure) shows subtle discrepancies in the strengths of other spectral lines in the
range of this plot too (right panel).}
\end{figure}

\subsection{Outlook}

Both results, the possible $\omega$~Cen collision event debris candidates and the COG's, need to be substantiated 
by new data. Especially the abundances available in the literature often do not cover all stars, and important species
are missing altogether. This especially applies to the $\omega$~Cen topic. Therefore, we have started gathering a more
uniform set of data, using FEROS at ESO-La Silla, FOCES at Calar Alto observatory in Spain, and SARG located 
at the TNG on the Canary Islands. These data are now in the process of reduction and analysis, and results should be
available in the course of 2008 (see Fig. 4). These new data will be able to verify whether any star in our sample can be  
properly characterised as $\omega$~Cen debris, and to further solidify our evidence for the COG group. 

Furthermore, the \citet{Peterson2001} objects (see Fig. 4) should be re-observed and accurate abundances obtained, so that the
relationship of the overdensity at ${\rm [Fe/H]}=-1.5$~dex to our COG's can be put on a firmer basis (or perhaps 
negated). A later step would be to extend the study to stars beyond the local regime. 

\section{The Impact of Gaia on Chemodynamics}
In the current pre Gaia era, an article on kinematics and abundances cannot conclude without at least
mentioning Gaia. Gaia will revolutionise our knowledge of our galaxy. This very ambitious astrometric satellite mission
will not only provide parallaxes and proper motions and photometry for $10^9$ stars, it will also derive abundances
and radial velocities for brighter subsets of the whole sample.\footnote{Actually Gaia's sample is not exactly a sample but
rather an {\em inventory} of all objects brighter than $V=20$~mag.} Hence relatively ``expensive'' studies like the one
described in this paper, using astrometry, photometry and high-resolution spectroscopy, will become a lot easier with 
Gaia data being available. The only ingredient that Gaia will not be able to deliver is detailed differential 
abundances (and radial velocities for the fainter stars), since its high-resolution spectrograph only covers 
a small spectral range. These will then need to be obtained by large ground-based
observatories utilising high-resolution spectrographs, possibly even high-resolution, multi-object spectrographs.
However, many slightly less ambitious studies can be done on the basis of Gaia data alone. 

By the time the Gaia results become available, the current ongoing large surveys, such as SDSS, Pan-STARRS, etc., 
will have accumulated a wealth of complementary data. This means that the amount of data readily available for 
chemodynamical and other studies of Galactic astronomy will in the next decade expand by several orders of 
magnitude, both in quantity {\em and} quality!  

All of these new developments will certainly contribute to a revolution in the way we see and understand our
Galaxy, and hence galaxies in general. At present we can confidently state that Galactic astronomers are
witnessing the dawn of very exciting times.

\acknowledgements 
Support for MC is provided by Proyecto Fondecyt Regular \#1071002. The authors would like to thank
the OPCs of Calar Alto, ESO and TNG for generously granting observing time and also M. Bellazzini 
for helping with the TNG/SARG proposal. MA expresses his thanks to the organisers of the sdOB conference for
the travel grant.

\end{document}